\pdfoutput=1
\documentclass{sigchi-ext}
\usepackage[T1]{fontenc}
\usepackage{textcomp}
\usepackage[scaled=.92]{helvet} 
\usepackage{graphicx} 
\usepackage{balance}  
\usepackage{booktabs} 
\usepackage{ccicons}  
\usepackage{ragged2e} 

\def\plaintitle{Explainable Interfaces for Rapid Gaze-Based Interactions in Mixed Reality
  Caps} 
\def\emptyauthor{}
\def\plainkeywords{Explainable AI; Mixed Reality, Eye Tracking}

\title{Explainable Interfaces for Rapid Gaze-Based Interactions in Mixed Reality
}

\numberofauthors{6}
\author{%
  \alignauthor{%
    \textbf{Mengjie Yu}\footnotemark[1]\\
    \textbf{Dustin Harris}\footnotemark[1]\\
    \textbf{Ian Jones}\\
    \textbf{Ting Zhang}\\
    \textbf{Yue Liu}\\
    \textbf{Naveen Sendhilnathan}\\
    \textbf{Narine Kokhlikyan}\\
    \textbf{Fulton Wang}\\
    \textbf{Co Tran}\\
    \textbf{Jordan L. Livingston}\\
    \textbf{Krista E. Taylor}\\
    \textbf{Zhenhong Hu}\\
    \textbf{Mary Anne Hood}\\
    \textbf{Hrvoje Benko}\\
    \textbf{Tanya R. Jonker}\\
    \affaddr{Meta Reality Labs Research}\\
    \affaddr{Redmond, Washington, US}\\
    \email{annaymj@meta.com} \\
} 
}
\definecolor{linkColor}{RGB}{6,125,233}
\hypersetup{%
  pdftitle={\plaintitle},
  pdfauthor={\emptyauthor},
  pdfkeywords={\plainkeywords},
  bookmarksnumbered,
  pdfstartview={FitH},
  colorlinks,
  citecolor=black,
  filecolor=black,
  linkcolor=black,
  urlcolor=linkColor,
  breaklinks=true,
}

\begin{document}

\CopyrightYear{2020}
\setcopyright{rightsretained}
\conferenceinfo{CHI'24 HCXAI Workshop}{May  11--16, 2024, Honolulu, HI, USA}
\isbn{978-1-4503-6819-3/20/04}
\doi{https://doi.org/10.1145/3334480.XXXXXXX}
\copyrightinfo{\acmcopyright}

\maketitle

\RaggedRight{} 

\begin{abstract}
Gaze-based interactions offer a potential way for users to naturally engage with mixed reality (XR) interfaces. Black-box machine learning models enabled higher accuracy for gaze-based interactions. However, due to the black-box nature of the model, users might not be able to understand and effectively adapt their gaze behaviour to achieve high-quality interaction. We posit that explainable AI (XAI) techniques can facilitate understanding of and interaction with gaze-based model-driven system in XR. To study this, we built a real-time, multi-level XAI interface for gaze-based interaction using a deep learning model, and evaluated it during a visual search task in XR. A between-subjects study revealed that participants who interacted with XAI made more accurate selections compared to those who did not use the XAI system (i.e., F1 score increase of 10.8\%). Additionally, participants who used the XAI system adapted their gaze behavior over time to make more effective selections. These findings suggest that XAI can potentially be used to assist users in more effective collaboration with model-driven interactions in XR.

    \footnotetext[1]{The two authors contributed  equally to this paper.}

\end{abstract}




\section{Introduction}
Interacting with mixed reality interfaces using gaze has emerged as a promising interaction modality, as it can offer a fast and intuitive method for targeting and selection. In prior work, gaze has been used in a direct manner for pointing\cite{Rajanna2018} and/or selection\cite{Mohamed2018} of user interface components. However, more recently, research has paid increasing attention to model-based approaches that leverage natural gaze patterns \cite{david_john2021}\cite{peacock2022gaze}\cite{sendhilnathan2022detecting}\cite{pfeuffer2023palmgazer} and scanpaths \cite{burlingham2024motor}. In these instances, users use their natural eye movement implicitly, in a goal-directed manner \cite{sendhilnathan2021neural} and user's gaze behavior is used to infer intentions, cognitive load \cite{gupta2023investigating} and skill levels\cite{Jacob2019}\cite{Bhanuka2022}. These gaze-based interaction inferences are often driven by black-box machine learning models, which tend to achieve higher prediction accuracy. Although these models can make more accurate predictions, they are black-boxes, meaning they are difficult for users to understand errors and build a mental model for effective use. In other words, learning gaze-based interactions can be challenging for users.

\begin{marginfigure}[-35pc]
  \begin{minipage}{\marginparwidth}
    \centering
    \includegraphics[width=0.9\marginparwidth]{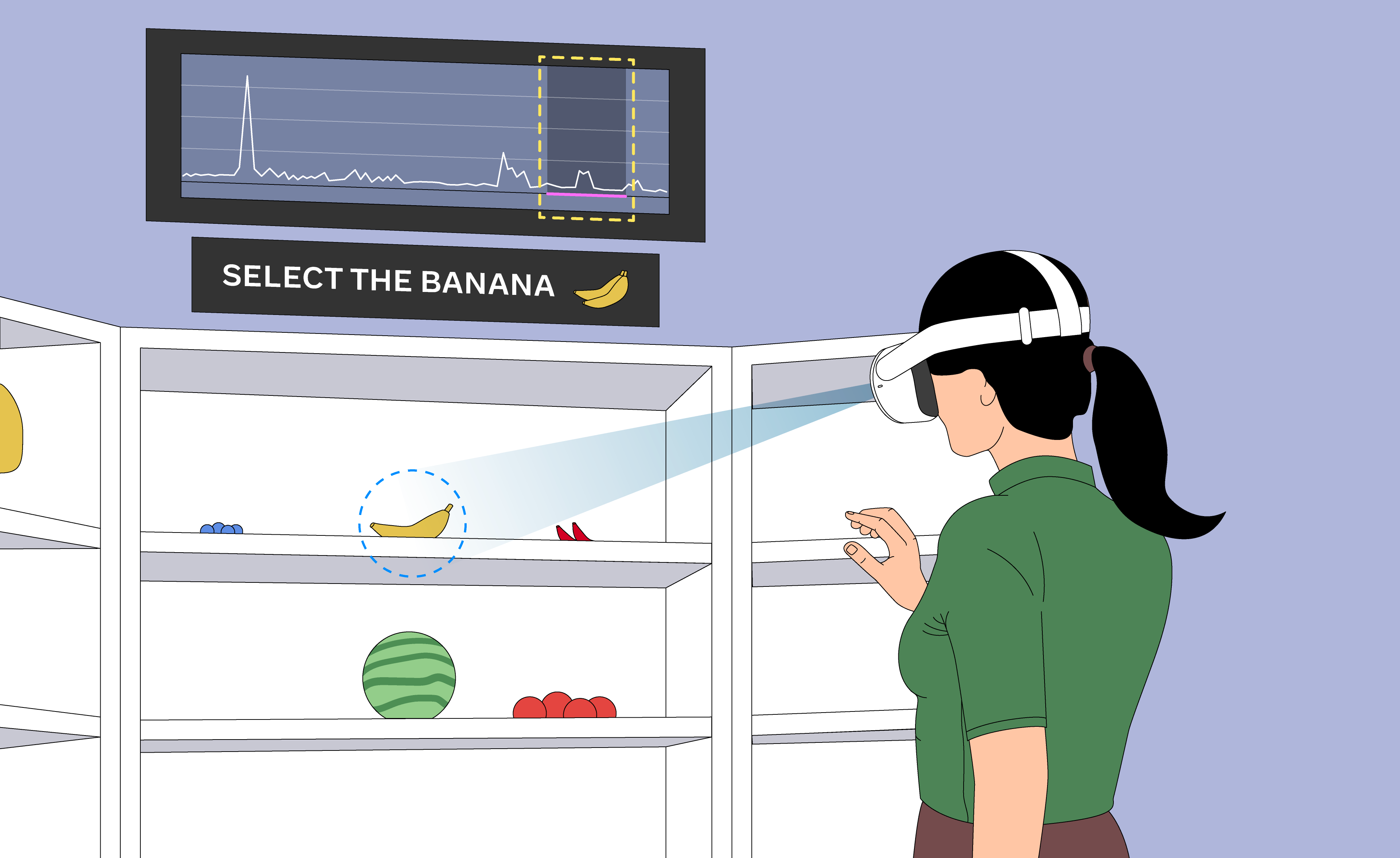}
    \caption{Visualization of the explainable interface designed to aid users in understanding gaze-based interactions driven by a model in a mixed reality setting. The dashed blue circle indicates the object being gazed upon, while the graph above the panel displays the user’s eye gaze patterns over time. The dashed pink line within the yellow rectangle represents a counterfactual explanation, guiding the user on how to modify their eye gaze behavior to adjust the model’s output.}~\label{figure_1}
  \end{minipage}
\end{marginfigure}
Designing an XAI interface for rapid interaction using real-time sensor data in XR remains an unsolved problem. Prior work showed that integrating XAI in user interface design can enhance transparency and interpretability of AI systems\cite{miller2019explanation}, aid users in understanding AI predictions\cite{ribeiro2016should}, and facilitate user's comprehension of complex AI models\cite{samek2021explaining}. Recent work proposed a framework of when, what and how to show explanations for model output in everyday augmented reality\cite{xu2023xair}. However, XAI for sensing and interaction is the relatively under-explored area. We posit that XAI can serve as a bridge to help users better learn and understand AI-powered interaction system, thus further enabling users to interact more efficiently with AI-powered system in XR.

To study this, we developed an adaptive interface that simulates an AR interface using a real-world scene. This interface uses a gaze model to automatically highlight a target of interest and provide a real-time, multi-level XAI interface that explains the model's behaviour. We conducted a user study as an initial step to answer the following research questions:  
\begin{itemize}
    \item \textbf{RQ1:} Can XAI help users understand model-driven gaze-based interaction in XR?
    \item \textbf{RQ2:} What types of explanations most effectively enhance understanding during rapid interaction in XR?  
\end{itemize}

\begin{figure*}[!h]
  \centering
  \includegraphics[width=2.2\columnwidth]{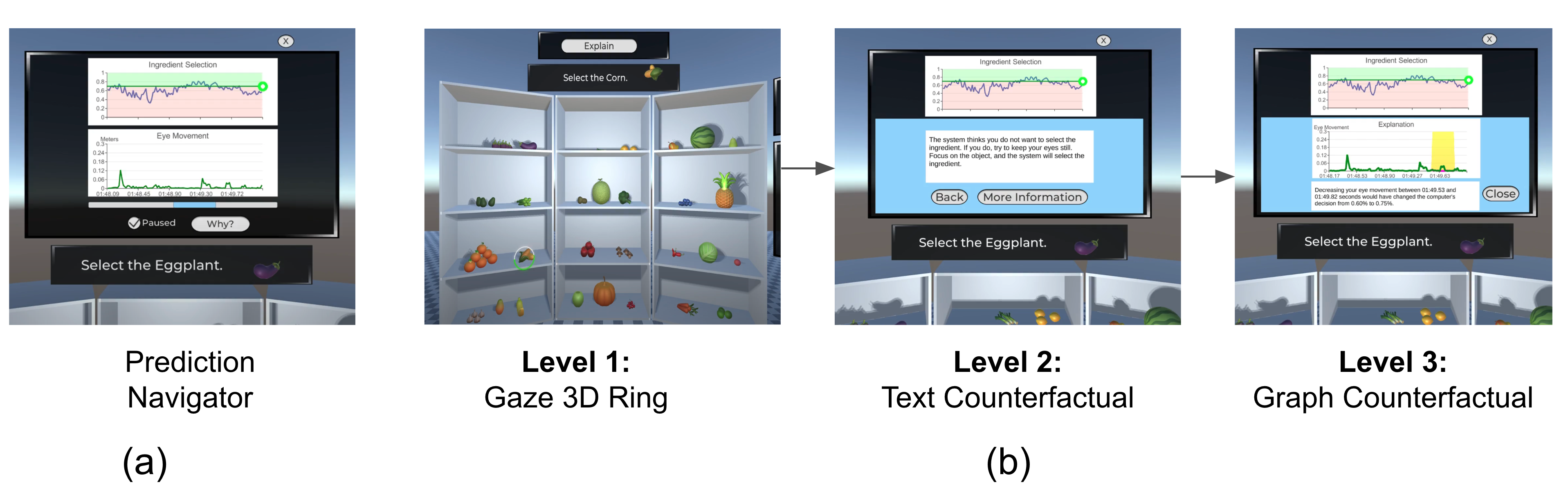}
  \caption{ Screenshots from the prototype showing the prediction navigator (a) and 3 Levels of Explanation (b) }
  \label{figure_2}
\end{figure*}

\section{Method}
\begin{figure*}[!h]
  \centering
  \includegraphics[width=2\columnwidth]{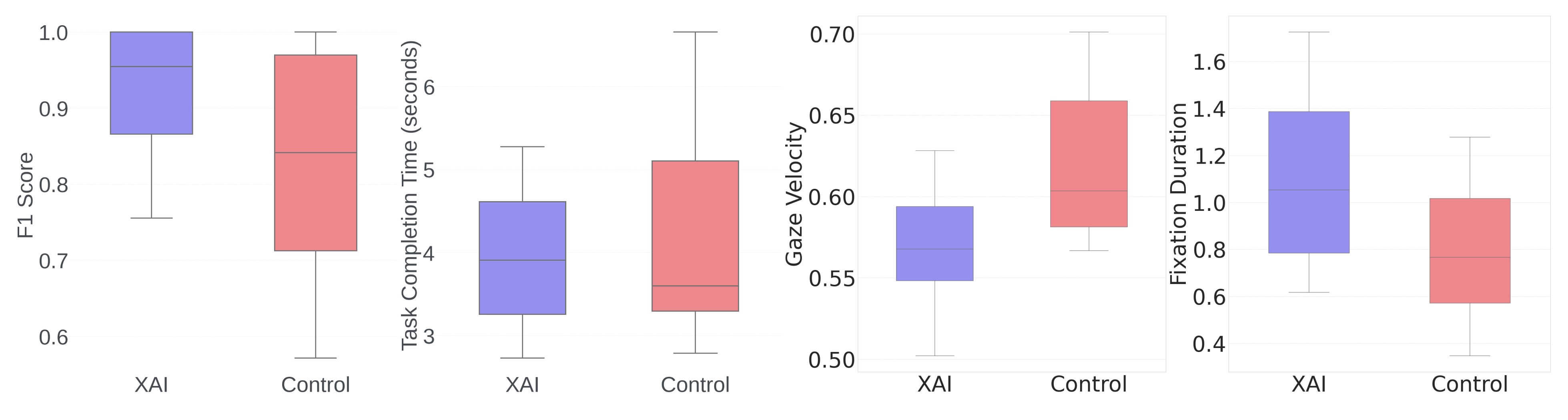}
  \caption{Boxplot distribution of task performance and gaze behavior metrics between the XAI and control conditions. From left to right, F1 score, task completion time, gaze velocity and fixation duration, respectively. }
  \label{figure_3}
\end{figure*}
We developed a temporal convolutional network (TCN) model \cite{tcn} that predicts the probability of target selection, which was trained using a gaze dataset collected in a prior study \cite{david_john2021}. To generate explanations, we used SHAP counterfactual method \cite{Ramon_2020}, which visualizes the minimal change needed to flip the model's prediction. 

A between-subjects study was conducted with 32 participants, in which users were asked to locate and select a specific target. Participants completed selection tasks in a virtual environment on a Meta Quest 2 with a custom-built Tobii Crystal eye tracker. The study consisted of pre-trial, training, and testing phases, with each phase containing 5 tasks.  Gaze logging and user feedback were collected. Three levels of explanations were offered in a progressive disclosure manner, ranging from a high-level visualization of the model’s output probability (i.e., Level 1: Gaze 3D Ring) to detailed information about how the model uses gaze data to predict the user’s intent to interact with the VR environment in textual (Level 2: Text counterfactual) and graphical (Level 3: Graph counterfactual) format(See Figure \ref{figure_2}).  

\section{Results}
Above we show preliminary results from our study including gaze motion metrics and insights from participants’ qualitative feedback. The leftmost panel, we show how XAI benefits users on task performance. F1 score and task completion time during the testing period were analyzed and compared using paired-samples t-tests. The results showed that the average F1 score in XAI condition was significantly ($t(29)=2.206; p < 0.05$) higher ($M = 0.92, SD = 0.09$)  than the control condition $(M = 0.83, SD = 0.14)$ in Figure \ref{figure_3}. However, no significant differences were detected in task performance time. The results suggest users made selections based on their gaze more precisely, but this could potentially increase the time it takes to complete the task. 

We also evaluated the eye-tracking data collected during the task. To examine the behavior of the gaze dynamics, we compared the participants’ gaze velocity and fixation duration between the two conditions. Participants in the XAI condition ($M = 0.57, SD = 0.036$) exhibited lower gaze velocity ($t(29)= 3.13, p < 0.05$) compared to those in the control group  ($M = 0.62, SD = 0.048$), suggesting that the XAI group had more nuanced and controlled gaze behavior, thus supporting our hypothesis that XAI interfaces facilitate a better understanding of rapid interaction models in XR. Additionally, participants in the XAI condition ($M = 1.07, SD = 0.36$) had significantly higher fixation duration values ($t(29) = 2.17, p< 0.05$) compared to the control group ($M = 0.79, SD = 0.29$) suggesting that these participants had a learned proficiency in gaze control, thus enabling them to concentrate more effectively on task-related objects, thereby enhancing their selection capabilities. Collectively, these findings underscore the impact of XAI interfaces in influencing and enhancing users’ gaze behaviors in the context of XR interaction models.

In order to gain more insight into preferred explanation types, participants were asked open-ended questions about their likes, dislikes, and suggestions regarding the provided explanations. The coded responses resulted in a $.95$ total percent agreement and a Cohen’s kappa value of $.81$ across coders and themes. Seven themes emerged on 'what’ and ‘how’ to show  XAI for rapid interaction in XR. Users expressed a desire for \textbf{real-time} explanations during the interaction and \textbf{improved tutorials} before starting to use the system. We found that users want \textbf{low complexity} in explanations, but also \textbf{control over the display of information} to alter the level of complexity based on their needs, even showing \textbf{multiple explanations} where each provides unique insight into the system. When using the system, users prefer to learn through \textbf{reinforced feedback}, where the system notifies them of how well they are performing on the task. Finally, these explanations should \textbf{map to the behavior} of the user, for example, showing the user how they might steady their gaze to alter the system outcome.

\section{Conclusion and Discussion }
Our work contributes to the field of XAI in XR by exploring the value of XAI and user preferences of explainable interface for eye-tracking based interaction system. We hypothesize that XAI can serve as a bridge for user to understand and collaborate more efficiently with AI powered interaction system. Our preliminary results showed significant improvement in task performance indicated by model performance (e.g., F1 score) and users’ adapted eye gaze behavior (e.g., gaze velocity and fixation duration) with XAI. Our qualitative analysis summarized 7 themes of user preferences for XAI that can help guide future explainable interface design in XR. 

\balance{} 

\bibliographystyle{SIGCHI-Reference-Format}
\bibliography{sample}

\end{document}